\documentclass[12pt]{article}
\usepackage{epsfig}

\newcommand{\be}{\begin{eqnarray}}
\newcommand{\ee}{\end{eqnarray}}
\newcommand{\sll}{\raise.15ex\hbox{$/$}\kern-.43em\hbox{$l$}}
\newcommand{\slp}{\raise.15ex\hbox{$/$}\kern-.43em\hbox{$p$}}
\newcommand{\slq}{\raise.15ex\hbox{$/$}\kern-.43em\hbox{$q$}}
\newcommand{\slk}{\raise.15ex\hbox{$/$}\kern-.43em\hbox{$k$}}
\newcommand{\slepsilon}{\raise.15ex\hbox{$/$}\kern-.53em\hbox{$\epsilon$}}

\begin{document}

\bibliographystyle{unsrt}
\footskip 1.0cm

\thispagestyle{empty}
\vspace{1in}

\begin{center}{\Large \bf {Enhancement and Suppression of the 
Neutrino-Nucleon Total Cross Section at Ultra-High Energies}}\\

\vspace{1in}
{\large  Jamal Jalilian-Marian}\\

\vspace{.2in}
{\it Physics Department, Brookhaven National Laboratory,
Upton NY 11973\\ }

\end{center}

\vspace*{25mm}

\begin{abstract}

\noindent We argue that high gluon density effects at small $x$ are  
important for the calculation of ultra-high energy neutrino nucleon cross 
sections due to the phenomenon of geometric scaling. We calculate the 
cross section for $\nu \, N \rightarrow \mu \, X$, including high gluon 
density effects, using the all twist formalism of McLerran and 
Venugopalan and show that it can be related to the dipole nucleon 
cross section measured in DIS experiments. For neutrino energies of 
$E_{\nu}\sim 10^{12}$ GeV, the geometric scaling region extends all 
the way up to $Q^2 \sim M^2_{W}$. We show that geometric scaling 
can lead to an {\it enhancement} of the neutrino nucleon total cross section
by an order of magnitude compared to the leading twist cross section 
and discuss the implications for neutrino observatories. At extremely 
high energies, gluon saturation effects suppress the neutrino nucleon 
total cross section and lead to its unitarization.

\end{abstract}
\newpage

\section{Introduction}

Ultra high energy neutrinos are a source of great mystery and excitement
and offer a possible window onto phenomena beyond the standard model.
Because of their weak interactions with matter, neutrinos can travel large 
distances and therefore carry information about very distant objects.
The origins of ultra high energy neutrinos are uncertain and the subject 
of intense theoretical and experimental interest and investigation \cite{fh}.
Some possible sources are active galactic nuclei, decays of super heavy
particles and gamma ray bursts. They can be detected by measuring the
air showers initiated by the muon produced in neutrino-nucleon interactions 
through a charged current.
 
The total neutrino nucleon cross section can be calculated 
\cite{ptqcd} in the standard model using the various parametrizations 
of parton distribution functions \cite{parton} measured at HERA 
\cite{hera}. At very high neutrino energies, one needs to know the 
behavior of the parton distribution functions at small $x$. It can be 
shown that a power growth of the parton distribution functions with $x$ 
would lead to a power growth of the neutrino nucleon total cross section 
with neutrino energy. This would eventually lead to the violation of 
unitarity at high energies. 

Unitarization of ultra high energy neutrino nucleon cross sections has
been of considerable interest lately \cite{dicus,rsssv,stasto,gkr}. 
Saturation of the gluon distribution function at very small $x$ \cite{glr} 
is expected to restore unitarity at high energies. One can make a rough 
estimate of the magnitude of unitarity corrections at a given neutrino 
energy by considering the first higher twist correction factor 
$\alpha_s xG(x,Q^2)/ \pi R^2 Q^2$. Since the neutrino nucleon cross
section is dominated by scales $Q^2 \simeq M_W^2$, the effective value
of $x$ is $\sim {M_W^2 \over 2 M_h\,E_{\nu}}$. At a neutrino energy of
$E_{\nu} \sim 10^{12} $GeV, this is an effect of only a few percent. 
However, due to the phenomenon of geometric scaling, it is too naive to 
conclude that higher twist (high gluon density) effects can be 
disregarded. 

It is an experimental fact that the HERA data at small $x$ ($< 0.01$) and
all $Q^2$ show geometric scaling \cite{gscale}. In other words, the DIS
cross section depends on only one variable, $Q^2/Q^2_s (x)$ rather than
two independent variables $x$ and $Q^2$. Here $Q^2_s(x)$ is the saturation
scale of the nucleon, arising from high gluon density effects, which can 
be extracted from the HERA data \cite{gbw}. Geometric scaling is a property 
and prediction of the all twist formulation of QCD evolution equations for 
DIS structure functions and cross sections at small $x$ (high energy)
in the kinematic region $Q^2 < Q_s^2$ \cite{nonlin,bal}. Recently, it has 
been shown that the nonlinear evolution equations for the structure 
functions at small $x$ exhibit this geometric scaling property \cite{iim} 
(see also \cite{ks}) in the kinematic region beyond $Q_s^2$. Roughly, 
this means that high gluon density effects, which are dominant at scales 
$Q^2 \leq Q^2_s$, influence observables at much higher scales 
$Q^2 \gg Q_s^2$. This was used in \cite{klm} to fit the RHIC data on pion 
spectra at $p_t^2 \gg Q_s^2$ and to reproduce the $N_{part}$ scaling of 
the data. 

\begin{figure}[htp]
\centering
\setlength{\epsfxsize=12cm}
\centerline{\epsffile{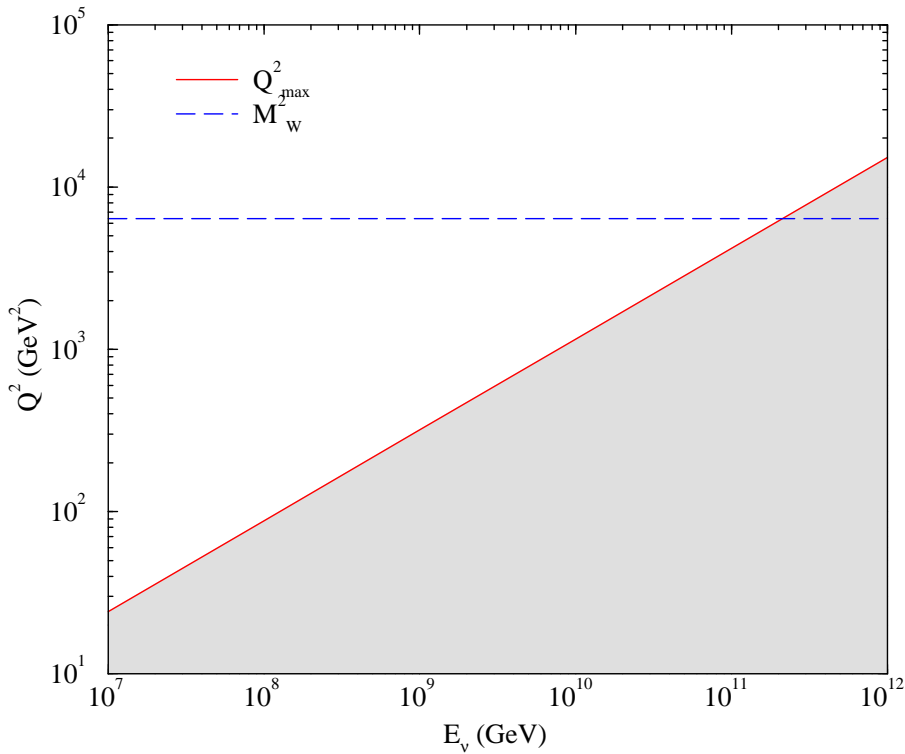}}
\caption{The geometric scaling region.}
\label{fig:Q_max}
\end{figure}

In \cite{iim} the $Q^2$ region where geometric scaling holds is calculated 
to be 
\be
Q^2_{max} \ll \bigg[ {Q_s^2(x) \over \Lambda_{QCD}^2}\bigg]\, Q_s^2(x)
\label{eq:gsregion}
\ee
The geometric scaling region for different neutrino energies is shown in 
Figure (\ref{fig:Q_max}). For easy reference, we also show $M_W^2$. We 
have used the Golec-Biernat and W\"usthoff parametrization of the 
saturation scale such that
\be
Q_s^2(x)\equiv Q_{s0}^2\,(x_0/x)^{\lambda}
\label{eq:gbw}
\ee
where $Q^2_{s0}=1.0$  $GeV^2$, $x_0=3.0 \times 10^{-4}$ and $\lambda = 0.28$,
$\Lambda_{QCD}=0.2$ GeV. Furthermore, since most of the contribution to 
the cross section comes from $Q^2 \sim M^2_W$, we have set  
$x= {M_W^2 \over 2M_h E_{\nu}}$.

Clearly, by $E_{\nu} \sim 10^{12} $ GeV we are in the geometric scaling 
region. Therefore, high gluon density and gluon saturation effects 
may be crucial. Since the standard expressions for the neutrino nucleon
cross sections are calculated within the leading twist perturbative QCD
formalism, they will break down due to the higher twist nature of gluon
saturation. Here, we use the effective action and renormalization group
approach to high energy QCD \cite{nonlin} to calculate this cross section 
including all higher twist (high gluon density) effects.

\section{Neutrino nucleon cross section}

In leading twist perturbative QCD, the expression for the neutrino nucleon 
differential cross section is given by 

\be
{d^2 \sigma^{\nu N} \over dxdQ^2}={G_F^2 \over \pi}
\bigg({M^2_{W,Z} \over Q^2 + M^2_{W,Z}}\bigg)^2 
\bigg[q(x,Q^2) + (1- Q^2/xs)^2 {\bar q}(x,Q^2)\bigg]
\label{eq:difcs}
\ee
where $x$ and $Q^2$ are the standard DIS variables and
the quark and anti-quark distributions include the appropriate 
couplings for neutral and charged currents in DIS. 
To get the total cross section, one integrates over $x$ and $Q^2$
\be
\sigma_{total}^{\nu N}(s)=\int_0^1 dx \int_0^{xs} dQ^2
{d^2 \sigma^{\nu N} \over dxdQ^2}
\label{eq:stcs}
\ee
At high neutrino energies and for very high values of $Q^2 >> M^2_{W,Z}$, 
the integrand dies off quickly while shrinkage of the phase space kills the 
contribution from the low momentum ($Q^2 << M^2_{W,Z}$) region. Therefore, the 
dominant contribution to the total cross section comes from the region
of $Q^2 \sim M_{W,Z}^2$ if one uses the standard parton distribution functions 
(by standard, we mean any of the available parametrizations of parton 
distributions such as MRS, CETEQ, GRV \cite{parton}). 

At small $x$, higher twist effects become important. This means that
the standard parton distribution functions, defined as the
expectation values of certain two point operators get contributions
from higher twist operators. This spoils their interpretation as a
number density. One can still define and calculate physical observables 
such as the structure function $F_2$ \cite{mv} that are experimentally 
measured. However, one cannot relate the all twist structure functions 
to number distributions such as the standard gluon distribution function 
$xG$. 

\subsection{Charged current exchange}

Here we use the effective action and classical field approach to
high gluon density effects to calculate the cross section for
the neutrino nucleon charged current interaction
\be
\nu_{\mu}\, N \rightarrow \mu \, X
\label{process}
\ee
Since we will ignore all lepton masses, our results would also apply 
to electron and tau neutrino scattering. Also, in this
work, we will ignore the neutral current exchange but it is quite
similar to the process considered here. Our goal here is to derive
an analytic expression for the above cross section such that it
includes all higher twist effects that are expected to unitarize it. 

We start with writing the differential cross section in terms of
leptonic and hadronic tensors
\be
{d\sigma \over dx\, dQ^2} = {1 \over 4 \pi}{y \over x s}
{G_F^2 M_W^4 \over [Q^2 + M_W^2]^2} \,
L^{\mu\nu}(k_1,k_2)\,W_{\mu\nu}(q^2, P\cdot q)
\label{eq:cslw}
\ee
where $k_1$,$k_2$ are the incoming and outgoing lepton momenta, 
$P$ is the nucleon momentum while $q=k_1-k_2$ is the momentum
transfer ($Q^2=-q^2$). The leptonic tensor $L_{\mu\nu}(k_1,k_2)$ 
is standard and is not affected by high gluon density effects. It is 
\be
L^{\mu\nu}(k_1,k_2) \equiv 2\bigg[k_1^{\mu}\,k_2^{\nu} +  k_1^{\nu}\,k_2^{\mu} 
- g^{\mu\nu}\,k_1\cdot k_2 + i\,\epsilon^{\mu\nu\rho\sigma}\,k_{1\rho}\,
k_{2\sigma}\bigg]
\label{eq:lepten}
\ee 
The hadronic tensor $W_{\mu\nu}$ contains all the information about
the high gluon density effect in a hadron. It is defined as
\be
W_{\mu\nu}(q^2, P\cdot q) \equiv {1 \over 2\pi} Im \int d^4z \, e^{iqz} 
<P|T\,J_{\mu}^{\dagger}(z)\,J_{\nu}(0)|P>
\label{eq:hadten}
\ee
where $J_{\mu}\equiv \bar{u}\gamma_{\mu}(1+\gamma_5)d$ is the charged 
weak current. The all twist hadronic tensor for electron proton DIS 
with a photon exchange has already been evaluated in \cite{mv}. Our
calculation here is a straightforward generalization to $W$ exchange
relevant for the process considered here. Since we are working with
a classical background field and external sources of color charge denoted
by $\rho$, we will need to generalize (\ref{eq:hadten}). This was already 
done in \cite{mv} where the hadronic tensor is defined as
\be
W_{\mu\nu}\equiv {\sigma \over 2\pi} {P^+ \over M_h} \,Im\, \int dX^-\,
\int d^4z \, e^{iqz} 
<T J_{\mu}^{\dagger}(X^- + z/2) J_{\nu}(X^- - z/2)>_{\rho}
\label{eq:genhadten}
\ee
where 
\be
<T J_{\mu}^{\dagger}(x) J_{\nu}(y)>_{\rho} = Tr\, 
\gamma_{\mu} (1+\gamma_5) S_u(x,y)\gamma_{\nu} (1+\gamma_5) S_d(y,x)
\label{eq:trjj}
\ee
and $S_{u,d}(x,y)$ is the $u$ or $d$ quark propagator in the 
background of the classical color field in coordinate space while  
$\sigma$ is the target hadron transverse area and $P^+$ is the
large component of the hadron momentum. In this approach, to calculate
a physical quantity, one averages over color charges $\rho$ at the end 
\cite{nonlin}. Despite its appearance,
the hadronic tensor defined in (\ref{eq:genhadten}) is Lorenz covariant
as discussed in \cite{mv}. The propagator in the background field is given 
by\footnote{Since we are ignoring quark masses, we will
not distinguish between $u$ and $d$ quarks and we drop the flavor label from
here on.} \cite{mv}  
\be
S(x,y) = S_0 (x,y) &-& i \int d^4r\,\bigg\{\bigg[ \theta (x^-)\theta (-y^-)
[V^{\dagger}(r_t) -1] -  \theta (-x^-)\theta (y^-)[V(r_t) -1]\bigg]
\nonumber \\
&&
S_0(x-r)\gamma^- \delta (r^-) S_0(r-y)\bigg\}
\label{eq:atprop}
\ee
with the free fermion propagator given by
\be
S_0(x-y)\equiv - \int {d^4\, p \over (2\pi)^4} e^{ip(x-y)} 
{{\slp} \over p^2 -i\epsilon}\,.
\label{eq:freeprop}
\ee
$V(r_t)$ is a matrix in fundamental representation which includes
the infinitely many gluon exchanges between the quark and the hadron.
The propagator also has other pieces that involve $\theta$ functions
on the same side in $x^-$ and $y^-$. As shown in \cite{mv}, these pieces
are pure gauges and do not contribute to the cross section. Therefore,
they are not included here. Finally, we will have to color average
our results to get the physical cross sections. We will come back to 
this point later when we discuss dipole models.

Using eqs. (\ref{eq:trjj}) and (\ref{eq:atprop}) in (\ref{eq:genhadten})
and after some lengthy algebra, we get
\be
W_{\mu\nu}= {N_c \sigma\over 2\pi} {P^+\over M_h} \,Im\, 
\int {d^4p \over (2\pi)^4}
{d^4k \over (2\pi)^4} (2\pi)\delta(k^-)\, \tilde{\gamma}(x,k_t,b_t)
{M_{\mu\nu} \over p^2\, (p-k)^2\, (p-q)^2\, (p-q-k)^2}
\label{eq:wmunuint}
\ee
where 
\be
\tilde{\gamma}(x,k_t,b_t)\equiv \int d^2 r_t \, e^{i k_t\cdot r_t} \,
\gamma(x,r_t,b_t) 
\label{eq:gammaft}
\ee
with
\be
\gamma (x,r_t,b_t) \equiv {1\over N_c} Tr\,
\left[\left<
1 - V(b_t +\frac{r_t}{2}) - V^\dagger(b_t -\frac{r_t}{2}) + 
V(b_t +\frac{r_t}{2}) V^\dagger(b_t -\frac{r_t}{2})
\right>_\rho \right]
\label{eq:gammadef}
\ee
and 
\be
M_{\mu\nu}\equiv 2\,Tr (1-\gamma_5)\gamma_{\mu}
({\slp}-{\slq} -{\slk})
\gamma^- ({\slp}-{\slq}) \gamma_{\nu}{\slp}
\gamma^- ({\slp} -{\slk})
\label{eq:Mmunu}
\ee
The function $\gamma (x,r_t,b_t)$ is related to the (quark-antiquark) 
dipole-nucleon scattering amplitude in coordinate space. 

\begin{figure}[htp]
\centering
\setlength{\epsfxsize=10cm}
\centerline{\epsffile{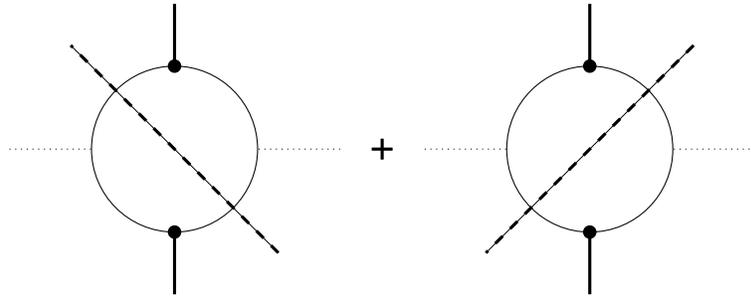}}
\caption{Imaginary part of the hadronic tensor.}
\label{fig:cut}
\end{figure}

To get the imaginary part of the hadronic tensor, we use the Landau-Cutkosky
cutting rules. There are two distinct ways of cutting the diagram as
shown in Fig. (\ref{fig:cut}). The dotted lines are $W$ boson gauge fields
while the thin solid lines represent fermions. The thick solid lines 
with a filled circle represent insertion of the classical field and 
the thick dashed lines represent the possible cuts. The cuts where 
both classical field insertions are on the same side are not kinematically 
allowed. 
The cut propagator is put on shell along with a theta function to ensure 
positive (negative) energy for fermions (antifermions). The sum of the 
two cuts is proportional to
\be
{\theta(p^+)\theta(q^+ - p^+ + k^+) \delta(p^2) 
\delta((p - k - q)^2) \over (p - k)^2\,(p - q)^2}
+
{\theta(p^+ - k^+)\theta(q^+ - p^+) \delta((p - k)^2) 
\delta((p - q)^2) \over p^2\,(p - k - q)^2} 
\nonumber
\ee
It is straightforward to show that the two contributions are actually
equal with appropriate change of variables and $\mu \leftrightarrow \nu$.
We get
\be
W_{\mu\nu}&=& {2N_c \sigma \over 2\pi}{P^+\over M_h} \,\int 
{d^4p \over (2\pi)^4}
{d^4k \over (2\pi)^4} (2\pi)\delta(k^-)\, \tilde{\gamma}(x,k_t,b_t)
\,M_{\mu\nu}
\nonumber \\
&\times&
{\theta(p^+ - k^+)\theta(q^+ - p^+) (2\pi)\delta((p - k)^2) 
(2\pi)\delta((p - q)^2) \over p^2\,(p - k - q)^2} 
\label{eq:wmunuint2}
\ee
where
\be
M_{\mu\nu} &\equiv& Tr \bigg[(1-\gamma_5)\gamma_{\mu}
({\slp}-{\slq} -{\slk})
\gamma^- ({\slp}-{\slq}) \gamma_{\nu}{\slp}
\gamma^- ({\slp} -{\slk}) \nonumber \\
&+&
(1-\gamma_5)\gamma_{\mu}
({\slp}-{\slq})
\gamma^- ({\slp}-{\slq}-{\slk}) \gamma_{\nu}({\slp} - {\slk})
\gamma^- {\slp}\bigg]
\label{eq:Mmunu1}
\ee
We can now use the delta functions to perform some of the integrals
in (\ref{eq:wmunuint2}). Defining $z\equiv p^-/q^-$, the effect of the
theta functions is to restrict the $z$ integration to the region between
$0$ and $1$. We then get
\be
M_h\,W_{\mu\nu} &=& {N_c\sigma \over 16\pi^2} 
{2P\cdot q \over q^- q^-} \int_0^1 dz \,
{d^2p_t \over (2\pi)^2}\,{d^2k_t \over (2\pi)^2}\,
\tilde{\gamma}(x,k_t,b_t)
\nonumber \\
&&
{\{[M^{sym}_{\mu\nu} + \mu \leftrightarrow \nu] + 
[M^{asym}_{\mu\nu} - \mu \leftrightarrow \nu]\} \over
[(p_t - zq_t)^2 -z(1-z)q^2][(p_t + k_t -zq_t)^2 -z(1-z)q^2]}
\label{eq:wmunuint3}
\ee
where $M^{sym}_{\mu\nu}$ and $M^{asym}_{\mu\nu}$ are now given by 
\be
M^{sym}_{\mu\nu} = Tr \, \gamma_{\mu}
({\slp} - {\slq} + {\slk})
\gamma^- ({\slp} - {\slq}) \gamma_{\nu}{\slp}
\gamma^- ({\slp} + {\slk})
\ee
and
\be
M^{asym}_{\mu\nu} = -\,Tr \, \gamma_5 \gamma_{\mu}
({\slp} - {\slq} + {\slk})
\gamma^- ({\slp} - {\slq}) \gamma_{\nu}{\slp}
\gamma^- ({\slp} + {\slk})
\ee
with $k^-=0$, $q^+=0$, $p^-=zq^-$ and 
\be
p^+ = - {(p_t - q_t)^2 \over 2(1-z)\,q^-}\;\;\;\;\;
\;\;\;\; k^+=-p^+ + {(p_t+k_t)^2 \over 2z\,q^-}
\ee
It is customary to write the hadronic tensor in terms of Lorenz invariant 
functions $W_1,W_2,W_3$ defined as \cite{rgr}
\be
M_hW_{\mu\nu}\equiv -(g_{\mu\nu} - {q_{\mu}q_{\nu} \over q^2})\,F_1 + 
{1 \over P\cdot q}(P_{\mu} - {q_{\mu}P\cdot q \over q^2})
(P_{\nu} - {q_{\nu}P\cdot q \over q^2})\,F_2 
+ i\,\epsilon_{\mu\nu\rho\sigma}{P^{\rho}q^{\sigma}\over P\cdot q}\,F_3
\label{eq:decomp}
\ee
where the structure functions are defined as $F_1=M_h\,W_1$, 
$F_2=\nu\, W_2$ and $F_3=\nu\, W_3$ with $M_h$ being the target nucleon 
mass and $P\cdot q = M_h \nu$. The differential cross section
$d\sigma/dxdQ^2$ can be written in terms of the structure functions
as
\be
{d\sigma \over dxdQ^2}= {1\over 2\pi}{G_F^2 \over x\,
[1 + {Q^2\over M_W^2}]^2}
\bigg\{y^2\, xF_1 + (1-y)\,F_2 + y[1-{y\over 2}]\,xF_3\bigg\}
\label{eq:dsigstruc}
\ee
which are related to the hadronic tensor via
\be
F_1 &=& -{1\over 2} \bigg[g^{\mu\nu} + 
{q^2 \over (q\cdot P)^2} P^{\mu}P^{\nu} \bigg] M_h W_{\mu\nu} \nonumber \\
F_2 &=& {q^2 \over 2 q\cdot P} \bigg[g^{\mu\nu} + 3 
{q^2 \over (q\cdot P)^2}  P^{\mu}P^{\nu} 
\bigg] M_h W_{\mu\nu}  \nonumber \\
&&
\nonumber\\
F_3 &=& \Pi^{\mu\nu}\, M_hW_{\mu\nu}
\label{eq:projec_f1f2f3}
\ee
where 
\be
\Pi^{\mu\nu} = -i\epsilon^{\mu\nu\alpha\beta}
{P_{\alpha}\, q_{\beta}\over 2 P\cdot q}
\ee
To give explicit expressions for the structure functions, we need to 
evaluate the traces\footnote{We would like to thank W. Vogelsang for his
help with evaluating these traces.}. This has already been done for the 
case of $W^{++}$ and $W^{--}$ in \cite{mv}. Using $K_0^{\prime} = -K_1$ and 
the identity
\be
\int_0^{\infty} dp \,{p\, J_0(p\,r_t) \over [p^2 + a^2]} \equiv K_0(a\,r_t)
\ee
gives

\be
2xF_1 &=& {N_c\, \sigma\, Q^2 \over 4\pi^3} \int^1_0 dz \int dr_t^2 \,
\gamma (x,r_t) \{ a^2 \,[z^2 + (1-z)^2]\, K_1^2 (ar_t)\}\nonumber\\
&&
\nonumber \\
F_2 &=& {N_c \sigma Q^2 \over 4\pi^3}\!\int^1_0 dz\!\int 
dr_t^2\,\gamma(x,r_t)
\bigg\{4 z^2(1-z)^2Q^2K_0^2(ar_t) + 
a^2 [z^2 + (1-z)^2]K_1^2(ar_t)\bigg\}\nonumber\\
&&
\nonumber \\
xF_3 &=& {N_c \sigma Q^2 \over 4\pi^3}\int^1_0 dz\int 
dr_t^2\, \gamma(x,r_t)
\bigg\{(1-2z)\,a^2 \,K_1^2(ar_t)\bigg\}
\label{eq:f1f2f3}
\ee
with $a^2\equiv z(1-z)Q^2$ and $K_0$ and $K_1$ are the modified Bessel 
functions. Using these expressions for the structure functions in 
(\ref{eq:dsigstruc}) and (\ref{eq:stcs}) gives the all twist cross 
section for $\nu \, N \rightarrow \mu \, X$. The structure functions
$F_1$ and $F_2$ are the same as those for electron nucleon deep inelastic
scattering while $F_3$ is specific to neutrino interactions. In the kinematics
of interest here (high energy or small $x$), $F_3$ vanishes as can be 
checked explicitly by doing the $z$ integration in (\ref{eq:f1f2f3}).   

One can distinguish three distinct kinematical regions in which the
neutrino nucleon total cross section has a different behavior. In the
very high energy limit where unitarity effects are dominant ($Q_s \ge M_W$), 
the cross section is given by (\ref{eq:f1f2f3}). This is the saturation 
region and the 
total cross section grows much more slowly (compared to the perturbative power
growth) due to high gluon density effects. At lower energies where 
$Q_s^2$ and $Q^2_{max}$, as defined in (\ref{eq:gsregion}), are both much 
less than $M_W^2$, one can use the standard perturbative results to which 
our expressions reduce\footnote{See \cite{mv} for a discussion of the 
high $Q^2$ limit of all twist calculations.}. 
The most interesting region is at high 
energies (but not too high where unitarity effects are dominant) where 
$Q^2_s \ll M_W^2$ but $Q^2_{max} > M_W^2$. Here, we are in the geometric 
scaling region where high gluon density effects lead to an enhancement 
of the unitarized cross section. We discuss this enhancement in the next 
section.

\section{Enhancement of neutrino-nucleon cross section}

\begin{figure}[htp]
\centering
\setlength{\epsfxsize=14cm}
\centerline{\epsffile{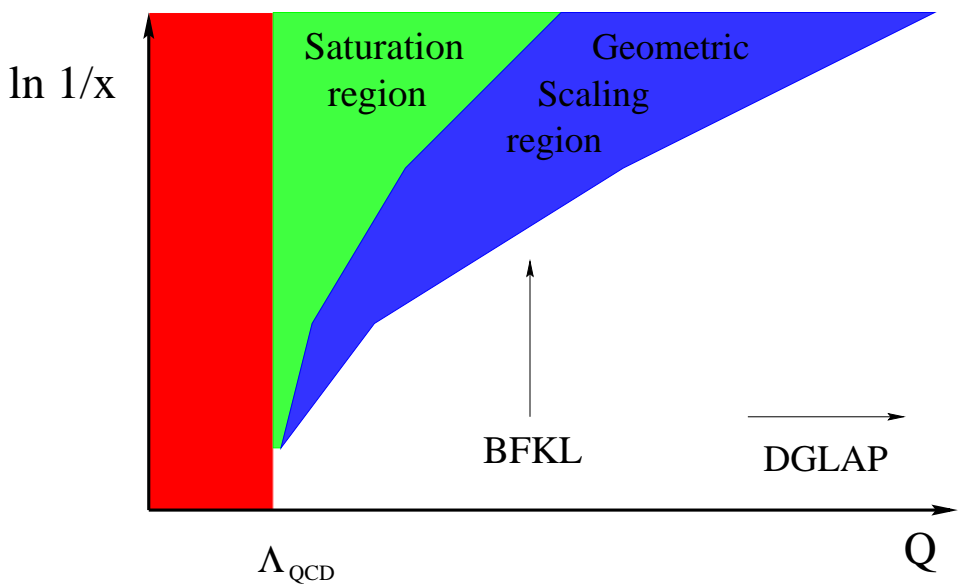}}
\caption{The saturation region.}
\label{fig:satreg}
\end{figure}

Geometric scaling was first observed at HERA for the virtual photon
nucleon total cross section \cite{gscale}. Roughly speaking, geometric 
scaling is the phenomenon that DIS structure functions depend on only
one variable $\tau\equiv Q^2/Q^2_s(x)$ rather than two independent
variables $x$ and $Q^2$. Geometric scaling arises naturally from the
all twist formulation of small $x$ QCD \cite{nonlin,bal}. It has been 
shown that geometric scaling is a property of the non-linear 
generalizations of the QCD evolution equation for the dipole cross section at 
small $x$ in the region where $Q^2 < Q_s^2$. More interestingly, 
it has been recently shown that the scaling region extends far beyond 
the saturation region \cite{iim,ks}, contrary to naive expectations
(see Fig. \ref{fig:satreg}). 
The dipole cross section is the universal building block which is 
also present in the unitarized all twist cross sections for particle 
production in $pA$ collisions \cite{pA}. 

To make an estimate of this enhancement factor for ultra high energy 
neutrinos, we need to know the dipole cross section $\gamma (x,r_t,b_t)$ 
beyond the classical approximation. In principle, one can determine 
$\gamma (x,r_t,b_t)$ from the renormalization group equations derived in 
\cite{nonlin}. In this approach, one solves the non-linear renormalization 
group equations on a lattice which can then be used to calculate the proton 
structure function $F_2$ which shows good agreement with the HERA data 
\cite{hw} at not very high $Q^2$. To improve the high $Q^2$ behavior, one
needs to include DGLAP evolution into the non-linear evolution equations. 
This is a very difficult problem which has not been solved yet. 
Alternatively, one can model the dipole 
cross section. One such model that includes DGLAP evolution is due to 
Bartels, Golec-Biernat and Kowalski \cite{gbw}. In this model, the 
dipole cross section is given by
\be
\gamma (x,r_t,b_t) = 
\bigg[1- exp[-\pi^2 \alpha_s \,r_t^2 \,xG(x,\mu^2)/3\sigma_0]\bigg]
\label{eq:dipole}
\ee
with $\mu^2= C/r_t^2 + \mu_0^2$ and $\sigma_0=23$ mb. This parametrization 
of the dipole cross section is used by \cite{gbw} to successfully fit 
all HERA data on inclusive and diffractive structure functions below 
$x=0.01$ and at all $Q^2$. However, this model will not have the 
enhancement since it still uses the DGLAP form of the dipole cross section
at small $r_t$ ($\gamma (x,r_t,b_t) \sim r_t^2 Q_s^2$).

One can also use the BFKL formalism to calculate the
dipole cross section in the geometric scaling region. This was
done in \cite{dino} where both LO and NLO BFKL evolution equations
were used and a resummation of most important collinear (and 
anticollinear) divergences was performed. In this formalism, the 
dipole cross section in the geometric scaling region is given by
\be
\gamma (x,r_t,b_t) =   
\bigg[r_t^2\, Q_s^2(x)\bigg]^{1-\gamma_s} 
\exp{\bigg\{- {1\over 2\beta \bar{\alpha_s} log 1/x}
\bigg[log[1/r_t^2\, Q_s^2(x)]\bigg]^2\bigg\}}
\label{eq:gamma}
\ee
where $\bar{\alpha_s}={N_c\alpha_s \over \pi}$ and $\beta=34$. A LO BFKL 
analysis leads to $1-\gamma_s=0.644$ while NLO BFKL leads to only a 
small change in this number. It is shown that the functional form of 
the saturation scale $Q_s^2$ depends on the energy considered. In the 
$x$ range of relevance here, however, it is a good approximation to use 
$Q_s^2(x)=\Lambda^2\, e^{\lambda_s log1/x}$ with $\lambda_s \sim 0.28$
and $\Lambda \sim 200$ MeV. 

A more desirable model would interpolate between the saturation
region and the geometric scaling region and have the DGLAP anomalous
dimension in the high $Q^2$ region. An ansatz for the dipole cross
section is given in \cite{gaussian} which interpolates between 
the saturation region and the geometric scaling region and fits the
HERA data at low $Q^2$ \cite{ei}. Work is in progress \cite{fj} to
improve this ansatz by including DGLAP evolution. It should be emphasized
that the leading twist high energy neutrino cross sections are dominated 
by a single scale $Q\sim M_W$ so that there is very little evolution in 
$Q^2$ while there is significant evolution in $x$ (by orders of magnitude!)
so that it is essential to treat the small $x$ evolution consistently.

In what follows, we will use the dipole cross section from the BFKL 
approach with the
effects of gluon saturation in the boundary between the saturation region
and the geometric scaling region taken into account, as given in 
(\ref{eq:gamma}), to make a rough numerical estimate of the enhancement 
factor. We will consider the ratio of the
dipole cross section in the coordinate space calculated from the
standard leading twist DGLAP evolution with anomalous 
dimension\footnote{The definition
of anomalous dimension is not standard. In our notation, the anomalous 
dimension is $1-\gamma_s$ so that the DGLAP anomalous dimension is $1$.}
of $1$ and the dipole cross section calculated from the BFKL approach with the 
saturation effects taken into account in the Geometric scaling region. 
The normalized dipole cross section (divided by a hadronic size 
$\sigma_0$ to make it dimensionless) is given by 
$[r_t^2 Q_s^2]$ in the standard leading twist approach 
and by $[r_t^2 Q_s^2]^{1-\gamma_0}$ in the BFKL approach (ignoring the 
exponential term which is very close to unity) with the high
gluon density effects taken into account in the boundary \cite{dino}.
Setting $r_t \sim 1/M_W$ since the cross section is dominated by $M_W$,
the ratio of cross sections in the geometric scaling region is
\be
\bigg[{M_W^2 \over Q_s^2(x)}\bigg]^{\gamma_s} \sim 10
\ee
with $Q_s^2 (x \sim 10^{-7} ) \sim 10 GeV^2$ and $\gamma_s=0.36$. One might 
think that this enhancement is due to the fact that BFKL cross sections 
grow faster than DGLAP cross sections. However, this is not the case. It 
is easy to 
show that the ratio of cross sections from the BFKL approach with and 
without geometric 
scaling is still larger than one (this ratio is about $3$, with choice of 
the infrared cutoff $k_0^2 = 1GeV^2$ \cite{stasto} in the BFKL apparoach 
without gluon
saturation effects in the boundary) so that the enhancement is due to the
geometric scaling. Since different values of $x$ contribute to the 
cross section, one will have different enhancement factors. This 
illustrates the fact that the neutrino nucleon cross section is 
enhanced in some neutrino energy range due to the geometric scaling property 
of the unitarized cross sections.

Off course, this is a very rough estimate but the enhancement should be
robust. In a quantitative analysis, one will have to include contributions 
from different regions of $x$ where one may or may not be in the geometric 
scaling region. At some neutrino energies, the dominant contribution will 
come from $x$'s where we will be fully in the geometric scaling region and 
the enhancement will be maximal\footnote{We are implicitly assuming that the
cross section is dominated by $Q \sim M_W$ even in the geometric scaling
region. It is possible that the effective $Q$ will shift to smaller values 
but this will make the enhancement even stronger.}. As one goes to yet 
higher energies, the geometric scaling region shrinks due to the fact that 
$Q_s^2\rightarrow M_W^2$ and one approaches the saturation region where 
unitarity effects become more important and eventually suppress 
the cross section compared to the leading twist cross section. A more 
quantitative analysis is in progress and will be reported elsewhere.

\section{Discussion}

We have calculated the total cross section for neutrino nucleon 
scattering via the charged current exchange including the high gluon 
density (higher twist) effects. We have shown that this cross section
is expressed in terms of the dipole nucleon cross section which
is the universal object appearing in all twist cross sections \cite{pA}. 
Using our expressions for the neutrino nucleon cross section and
some model of the dipole nucleon cross section (given for example in
\cite{gbw}), one can estimate at what neutrino energies protons will
look black to neutrinos (the black disk limit). This turns out to be 
at neutrino energies of $E_{\nu} \sim 10^{18}$ GeV. This may be too 
high of an energy for this effect to be observable in the near future. 

A more interesting effect happens at much smaller energies
than the black disk limit. As shown here, the geometric scaling region
extends all the way up to and beyond the weak boson mass already at
neutrino energies of $E_{\nu} > O(10^{12})$ GeV. Therefore 
neutrino nucleon cross sections at these energies will be dominated
by scales that are within the geometric scaling region where cross 
sections are typically enhanced. This enhancement factor can be as large 
as $1-2$ orders of magnitude at  $E_{\nu} > 10^{12}$ GeV. This will have 
very interesting consequences for neutrino astronomy and cosmology 
\cite{kus}.

This enhancement will be important for the current and future neutrino 
observatories \cite{kus}, \cite{agasa}, \cite{auger}, \cite{hires}, 
\cite{euso}, \cite{owl}. It will be very interesting to see whether the 
observed cosmic ray data points beyond the GZK cutoff \cite{gzk} are due to 
neutrinos \cite{weiler}. For 
these events to be neutrinos with their cross sections enhanced due to 
geometric scaling requires a very large enhancement factor already at 
$E_\nu \sim 10^{11-12}$ where the data points are \cite{agasa}. Several
experiments will detect neutrinos through their horizontal air showers
in the Earth's atmosphere due to neutrino air interactions whose rate 
will increase if the neutrino nucleon cross section is enhanced. On the
other hand, the rate of up-going air showers initiated by the leptons
produced in neutrino nucleon interactions will decrease if the neutrino
nucleon cross section is enhanced since the Earth will be less transparent
to neutrinos. A better understanding of the neutrino nucleon total cross
section is essential to these experiments which will help clarify the 
origins of ultrahigh energy neutrinos.

\leftline{\bf Acknowledgments} 

We would like to thank A. Dumitru, R. Harlander, K. Itakura, D. Kharzeev, 
W. Kilgore, S. Kretzer, A. Stasto, D. Teaney, D. Triantafyllopoulos, 
R. Venugopalan and W. Vogelsang for useful discussions. We would also like 
to thank L. McLerran for pointing out to us the importance of geometric 
scaling and its role in the enhancement of cross sections and many 
illuminating discussions. This work is supported by the U.S.\ Department 
of Energy under Contract No.\ DE-AC02-98CH10886 and in part by a PDF from 
BSA.

\leftline{\bf References}

\renewenvironment{thebibliography}[1]
        {\begin{list}{[$\,$\arabic{enumi}$\,$]}  
        {\usecounter{enumi}\setlength{\parsep}{0pt}
         \setlength{\itemsep}{0pt}  \renewcommand{\baselinestretch}{1.2}
         \settowidth
        {\labelwidth}{#1 ~ ~}\sloppy}}{\end{list}}

\end{document}